\documentstyle[aps,prl,floats]{revtex}
\input{epsf}
\begin{document}
\draft
\twocolumn[\hsize\textwidth\columnwidth\hsize\csname@twocolumnfalse\endcsname 
\title{Roles of repulsive and attractive forces in determining the structure of 
nonuniform liquids: generalized mean field theory} 
\author{John D. Weeks$^{1,2}$, Kirill Katsov$^1$, and Katharina Vollmayr$^1$} 
\address{$^1$Institute for Physical Science and Technology and\\ 
$^2$Department of Chemistry\\ 
University of Maryland, College Park, Maryland 20742} 
\date{\today } 
\maketitle 
 
\begin{abstract}
The structure of a nonuniform Lennard-Jones (LJ) liquid
near a hard wall is approximated by
that of a reference fluid with repulsive intermolecular forces
in a self-consistently determined external mean
field incorporating the effects of attractive forces.
We calculate the reference fluid structure by
first determining the response to the slowly varying part of the field
alone, followed by the response to the harshly repulsive part.
Both steps can be carried out very accurately, as confirmed by MC simulations,
and good agreement with the structure of the full LJ fluid is found.
\end{abstract} 
 
]

Dense liquids have highly nontrivial density correlations arising because
the harshly repulsive molecular cores cannot overlap \cite
{widomsci,wca,hansenmac}. Because of the constantly changing molecular
arrangements, such correlations play a much more fundamental role in liquids
than they do in other condensed phases such as glasses and solids, which
sample only a few basic configurations. Indeed, a model with only repulsive
intermolecular forces \cite{wca} can give a surprisingly accurate
description of the full density correlations seen in a uniform dense simple
liquid like Ar because the vector sum of the longer ranged attractive forces
on a given particle essentially {\em cancels} \cite{widomsci} in most
typical configurations.

{\em Nonuniform} liquids present a greater and qualitatively different
challenge, since even the averaged effects of attractive forces clearly do
not cancel \cite{wsb}. We discuss here an example where {\em both}
attractive and repulsive forces can greatly influence the liquid's
structure: a Lennard-Jones (LJ) fluid next to a hard wall. We obtain
accurate numerical results using a physically suggestive generalized mean
field description of the attractive forces \cite{ourotherpapers}. We
consider first the effects of these slowly varying forces on the liquid's
structure before taking account of the response to the rapidly varying (hard
core like) part of the external field. This treatment of attractive
interactions is quite different from that used in conventional integral
equation and density functional methods \cite{denfungeneral}, and we believe
it offers important conceptual and computational advantages.

Fluid particles interact with a known external (wall) field $\phi ({\bf r})$
and through the LJ pair potential $w(r_{ij})\equiv u_0(r_{ij})+u_1(r_{ij})$,
divided as usual \cite{wca} so that all the repulsive intermolecular forces
arise from $u_0$ and all the attractive forces from $u_1.$ We assume that
the external field $\phi ({\bf r})\equiv $ $\phi _0({\bf r})+\phi _1({\bf r}%
) $ can be separated in a similar way, where the subscript $0$ denotes in
all that follows a harshly repulsive interaction and the subscript $1$ a
much more slowly varying interaction usually associated with attractive
forces. Here we take $\phi ({\bf r})$ to be a hard wall potential, setting $%
\phi _1({\bf r})=0$ and $\phi _0({\bf r})=$ $\phi _{HW}(z),$ where $\phi
_{HW}=\infty $ for $z\leq 1$ (in reduced units) and $0$ otherwise, and we
let $\rho _B$ be the bulk density far from the wall.

We relate the structure of the nonuniform LJ\ system to that of a simpler 
{\em nonuniform} {\em reference fluid }\cite{wsb,wvk}, with only repulsive
intermolecular pair interactions $u_0(r_{ij})$ (equal to the LJ repulsions)
in a different {\em effective reference field} (ERF) $\phi _{\mbox{\tiny R}}(%
{\bf r})$. While the replacement of attractive pair interactions by an
approximate ``molecular field'' is an essential step in mean field theory,
we can think of other more general choices. Here we determine $\phi _{%
\mbox{\tiny R}}({\bf r})$ formally by the requirement that it has a
functional form such that the {\em local} (singlet) density at every point $%
{\bf r}$ in the reference fluid equals that of the full LJ fluid \cite
{sullstell}: 
\begin{equation}
\rho _0({\bf r;[}\phi _{\mbox{\tiny R}}])=\rho ({\bf r;[}\phi ])\,.
\label{singletden}
\end{equation}
The subscript $0$ reminds us that the reference system pair interactions
arise only from $u_0$ and the notation ${\bf [}\phi _{\mbox{\tiny R}}]$
indicates that all distribution functions are functionals of the appropriate
external field.

To find $\phi _{\mbox{\tiny R}}$ explicitly, we subtract the first equations
of the YBG hierarchy \cite{hansenmac} for the full and reference systems
with $\phi _{\mbox{\tiny 
R}}$ chosen so that Eq. (\ref{singletden}) is satisfied \cite{wsb,wvk}. The
result can be written exactly as 
\begin{eqnarray}
&-&\nabla _1[\phi _{\mbox{\tiny R}}({\bf r}_1)-\phi ({\bf r}_1)]=-\int d{\bf %
r}_2\rho _0({\bf r}_2|{\bf r}_1;[\phi _{\mbox{\tiny R}}])\nabla _1u_1(r_{12})
\nonumber \\
&&-\int d{\bf r}_2\{\rho ({\bf r}_2|{\bf r}_1;[\phi ])-\rho _0({\bf r}_2|%
{\bf r}_1;[\phi _{\mbox{\tiny R}}])\}\nabla _1w(r_{12})\,.  \label{exactybg}
\end{eqnarray}
Here $\rho _0({\bf r}_2|{\bf r}_1;{\bf [}\phi _{\mbox{\tiny R}}])\equiv \rho
_0^{(2)}({\bf r}_1,{\bf r}_2{\bf ;[}\phi _{\mbox{\tiny 
R}}])/\rho _0({\bf r}_1{\bf ;[}\phi _{\mbox{\tiny R}}])$ is the {\em %
conditional} singlet density, i.e., the density at ${\bf r}_2$ given that a
particle is fixed at ${\bf r}_1$.

If we assume that Eq. (\ref{singletden}) produces similar local environments
for the (identical) repulsive cores in the two fluids, which mainly
determine density correlations through excluded volume effects, then the
conditional singlet densities in the two fluids should also be very similar.
This key structural assumption introduces a generalized mean field theory in
which the reference fluid still has nontrivial pair and higher order
correlations induced by the repulsive forces. This suggests that the last
term on the R.H.S. in Eq. (\ref{exactybg}) is often very small. If we ignore
it completely \cite{wsbaccuracy} we obtain the approximate equation for the
field $\phi _{\mbox{\tiny R}}$ suggested by Weeks, Selinger, and Broughton 
\cite{wsb}: 
\begin{eqnarray}
\nabla _1[\phi _{\mbox{\tiny R}}({\bf r}_1)-\phi ({\bf r}_1)]=\int d{\bf r}%
_2\rho _0({\bf r}_2|{\bf r}_1;{\bf [}\phi _{\mbox{\tiny 
R}}])\nabla _1u_1(r_{12})\,.  \label{wsb}
\end{eqnarray}

Eq. (\ref{wsb}) incorporates mean field ideas, but it appropriately focuses
on {\em forces }\cite{widomsci,wca}. The relation to ordinary mean field
theory becomes clearer \cite{wvk} if we replace $\rho _0({\bf r}_2|{\bf r}_1;%
{\bf [}\phi _{\mbox{\tiny R}}])$ by $\rho _0({\bf r}_2;{\bf [}\phi _{%
\mbox{\tiny R}}])$ in Eq.\thinspace (\ref{wsb}). This approximation is much
better than one might at first suppose, since the main difference in these
two functions occurs when ${\bf r}_2$ is close to ${\bf r}_1,$ but then for
small $r_{12}$ the multiplicative factor $-\nabla _1u_1(r_{12})$ (the {\em %
attractive} part of the LJ force) vanishes identically. The gradient $\nabla
_1$ can then be taken outside the integral and Eq.\thinspace (\ref{wsb}) can
be integrated. Choosing the constant of integration so that the density far
from the wall equals $\rho _B$, we obtain the simplified mean field equation 
\cite{wvk}:

\begin{eqnarray}
\phi _{\mbox{\tiny R}}({\bf r}_{1})-\phi ({\bf r}_{1}) &\equiv &  \nonumber
\\
\phi _{s}({\bf r}_{1}) &=&\int d{\bf r}_{2}\,[\rho _{0}({\bf r}_{2};{\bf [}%
\phi _{\mbox{\tiny R}}])-\rho _{B}]\,u_{1}(r_{12})\,.  \label{mfint}
\end{eqnarray}

Because of the integration over the slowly varying attractive potential
``weighting function'' $u_1(r_{12})$, $\phi _s({\bf r})$ in Eq. (\ref{mfint}%
) extends smoothly into the repulsive core region where $\rho _0({\bf r};%
{\bf [}\phi _{\mbox{\tiny R}}])$ vanishes. Outside the wall it is smooth and
relatively slowly varying even when $\rho _0({\bf r};{\bf [}\phi _{%
\mbox{\tiny R}}])$ itself has pronounced oscillations. Physically $\phi _s(%
{\bf r})$ mimics the effects of the unbalanced attractive forces in the LJ
system, giving a soft {\em repulsive} interaction \cite{wsb} that tends to
push the reference particles away from the wall.

In order to solve equations like (\ref{wsb}) or (\ref{mfint}) to obtain the
self-consistent ERF $\phi _{\mbox{\tiny R}}({\bf r})$, we must determine the
required reference fluid distribution functions arising from a given
external field. In previous work \cite{wsb,wvk}, computer simulations were
used for this purpose. We now introduce a simple and accurate numerical
method for calculating these distribution functions and illustrate it here
by solving (\ref{mfint}) for the case of the LJ fluid near the hard wall.

We note that the ERF $\phi _{\mbox{\tiny 
R}}({\bf r})\equiv \phi _{\mbox{\tiny 
R}0}({\bf r})+\phi _{\mbox{\tiny 
R}1}({\bf r})$ in Eq. (\ref{mfint}) (and other related equations) can be
naturally separated into the sum of a harshly repulsive part, $\phi _{%
\mbox{\tiny 
R}0}({\bf r}),$ and a much more slowly varying ``smooth'' part $\phi _{%
\mbox{\tiny 
R}1}({\bf r}),$ arising physically mainly from the attractive interactions
in the original system. Eq. (\ref{mfint}) suggests the identification $\phi
_{\mbox{\tiny 
R}0}({\bf r})=$ $\phi _0({\bf r})$ and $\phi _{\mbox{\tiny 
R}1}({\bf r})=$ $\phi _s({\bf r})+\phi _1({\bf r}).$ More generally, we can
define $\phi _{\mbox{\tiny R}0}({\bf r})=$ $\phi _0({\bf r})-\phi _{0s}({\bf %
r})$ and $\phi _{\mbox{\tiny R}1}({\bf r})=$ $\phi _s({\bf r})+\phi _{0s}(%
{\bf r})+\phi _1({\bf r}),$ where $\phi _{0s}({\bf r})$ is an essentially
arbitrary smooth function that is nonzero only in the repulsive core region
but with $\phi _{0s}({\bf r})<<\phi _0({\bf r})$, so that $\phi _{%
\mbox{\tiny 
R}0}$ remains a harshly repulsive interaction. In the present case it is
sufficient to take the separation suggested by Eq. (\ref{mfint}), with $\phi
_{\mbox{\tiny 
R}0}=$ $\phi _{HW}$ and $\phi _{\mbox{\tiny 
R}1}=\phi _s$.

Our task is now to determine the local density $\rho _0({\bf r};{\bf [}\phi
_{\mbox{\tiny 
R}}])\equiv $ $\rho _{0,\mbox{\tiny 
R}}({\bf r})$ produced by a given ERF $\phi _{\mbox{\tiny 
R}}$. We provide a new way to solve this basic problem, quite independent of
its origins in the mean field equation (\ref{mfint}). Initially we treat the
LJ repulsive potential $u_0$ as a hard core interaction, but then use
standard methods \cite{hansenmac} to correct for its finite softness in our
final numerical results. We expect that there will be very different
responses of the reference fluid density to the rapidly and slowly varying
parts of the ERF $\phi _{\mbox{\tiny R}}\equiv \phi _{\mbox{\tiny R}0}+\phi
_{\mbox{\tiny
R}1}$ and anticipate that any large oscillations arise mainly from the
harshly repulsive part $\phi _{\mbox{\tiny R}0}$. These oscillations cause
problems in density functional methods, which use a variety of somewhat
arbitrary ``weighting functions'' to arrive at some underlying ``smooth
density'' for use in a free energy functional \cite{denfungeneral}.

\begin{figure}[tbp]
\epsfxsize=2.8in
\centerline{\epsfbox{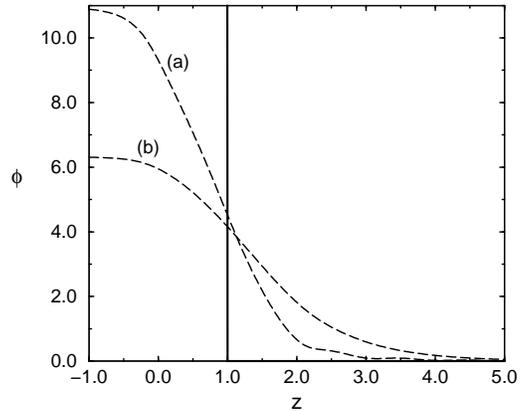}}
\caption{Self-consistent potential $\phi _{{\mbox{\tiny R}}1}=\phi _s$
(dashed line) for $\rho _B=0.785$ (a) and $\rho _B=0.45$ (b), $T=1.35$, and
bare wall potential $\phi _{{\mbox{\tiny R}}0}=\phi _{HW}$ (solid line). The
ERF $\phi _{\mbox{\tiny R}}=\phi _{{\mbox{\tiny R}}0}+\phi _{{\mbox{\tiny
R}%
}1}$.}
\end{figure}

Instead, we first determine the response to the {\em slowly varying part of
the ERF alone}, followed by the response to the harshly repulsive part,
using different methods in each step appropriate for the different density
responses. In the first step, we determine the associated ``smooth
interface'' $\rho _0({\bf r;[}\phi _{\mbox{\tiny 
R}1}])\equiv $ $\rho _{0,\mbox{\tiny 
R}1}({\bf r})$ that arises naturally from the slowly varying part $\phi _{%
\mbox{\tiny 
R}1}=\phi _s$ of the ERF {\em alone.} Physically, this takes account of the
effects of the attractive interactions modeled by $\phi _{\mbox{\tiny
R}1}$. We start from the basic equation relating small changes in the
potential and density \cite{hansenmac}: 
\begin{equation}
-\beta \delta \phi _{\mbox{\tiny 
R}1}({\bf r}_1)\,{\bf =}\int d{\bf r}_2\chi _0^{-1}({\bf r}_1,{\bf r}_2;{\bf %
[}\rho _{0,\mbox{\tiny 
R}1}])\delta \rho _{0,\mbox{\tiny 
R}1}({\bf r}_2)\,,  \label{linresp}
\end{equation}
through the generalized linear response function $\chi _0^{-1}({\bf r}_1,%
{\bf r}_2;{\bf [}\rho _{0,\mbox{\tiny 
R}1}])\equiv \delta ({\bf r}_1\!-\!{\bf r}_2)/\rho _{0,\mbox{\tiny 
R}1}({\bf r}_1)\!-\!c_0({\bf r}_1,{\bf r}_2;{\bf [}\rho _{0,\mbox{\tiny 
R}1}])$. Here $c_0({\bf r}_1,{\bf r}_2;{\bf [}\rho _{0,\mbox{\tiny 
R}1}])$ is the direct correlation function of the reference fluid with
density $\rho _{0,\mbox{\tiny R}1}$ and $\beta =1/k_BT.$ Specializing to the
case when the change is a small displacement of the field, we find the exact
equation \cite{lmb}: 
\begin{eqnarray}
\nabla _1\rho _{0,\mbox{\tiny 
R}1}({\bf r}_1) &/&\rho _{0,\mbox{\tiny 
R}1}({\bf r}_1)=-\beta \nabla _1\phi _{\mbox{\tiny R}1}({\bf r}_1)  \nonumber
\\
&+&\int d{\bf r}_2c_0({\bf r}_1,{\bf r}_2;{\bf [}\rho _{0,\mbox{\tiny R}%
1}])\nabla _2\rho _{0,\mbox{\tiny R}1}({\bf r}_2)\,.  \label{LMB}
\end{eqnarray}

If $\rho _{0,\mbox{\tiny 
R}1}$ is relatively slowly varying, we can accurately approximate $c_0({\bf r%
}_1,{\bf r}_2;{\bf [}\rho _{0,\mbox{\tiny 
R}1}])$ under the integral in Eq. (\ref{LMB}) by the {\em uniform fluid}
function $c_0(r_{12};\bar{\rho}_{12}),$ where $\bar{\rho}_{12}$ is some
intermediate density associated with the two points \cite{frishleb}. A
natural choice that gives very good results when $\rho _{0,\mbox{\tiny 
R}1}$ is reasonably smooth is $\bar{\rho}_{12}=[\rho _{0,\mbox{\tiny 
R}1}({\bf r}_1)+\rho _{0,\mbox{\tiny 
R}1}({\bf r}_2)]/2$. Starting with a given $\phi _{\mbox{\tiny 
R}1}$, we can then solve Eq. (\ref{LMB}) for the associated $\rho _{0,%
\mbox{\tiny 
R}1}$ by iteration, making use of the analytic and accurate Percus-Yevick
(PY) expressions for the direct correlation function of the uniform hard
sphere fluid \cite{frishleb,hansenmac}. If necessary, we can choose $\phi _{%
\mbox{\tiny 
R}1}$ inside the repulsive core region to help ensure that $\rho _{0,%
\mbox{\tiny 
R}1}$ is smooth enough for the expansion method to be accurate; this
procedure is important in some other applications \cite{ourotherpapers}.

A special case where this step can be carried out analytically arises when $%
\rho _{0,\mbox{\tiny R}1}$ varies so slowly that it is accurate to keep only
the first term in the expansion of $\nabla _2\rho _{0,\mbox{\tiny 
R}1}({\bf r}_2)\,$ in Eq. (\ref{LMB}) about ${\bf r}_1$. After integrating,
we arrive at the simple {\em local hydrostatic relation} \cite{frishleb}
between $\rho _{0,\mbox{\tiny 
R}1}$ and $\phi _{\mbox{\tiny 
R}1}$: 
\begin{equation}
\mu _0(\rho _{0,\mbox{\tiny 
R}1}({\bf r}))+\phi _{\mbox{\tiny 
R}1}({\bf r})=\mu _{0B}\,,  \label{hydrostatic}
\end{equation}
where $\mu _0(\rho )$ is the chemical potential of the uniform (hard sphere)
reference fluid at density $\rho $ and $\mu _{0B}=$ $\mu _0(\rho _B).$

The smooth profile $\rho _{0,\mbox{\tiny R}1}$ is analogous to one that
could be calculated using a single occupancy lattice gas (Ising) model,
where correlations arise {\em only} from attractive interactions \cite
{widomsci}. A realistic fluid has additional short wavelength correlations
due to the repulsive intermolecular forces. These show up primarily in the
second step of our method, where we take account of the response to $\phi _{%
\mbox{\tiny R}0}$, the remaining harshly repulsive part of the ERF.

Consider first a {\em small} perturbing potential $\delta \phi _{%
\mbox{\tiny
R}0}$ that is nonzero only inside the wall region with $z<1$. Evaluating Eq.
(\ref{linresp}) for $z_1>1$ gives an exact relation between the small
induced density changes inside and outside the wall region. However, it has
been shown that even large density fluctuations in a hard sphere fluid are
accurately described by gaussian fluctuation theory \cite{crooks}. This
suggests that if we could somehow {\em impose} the proper values on the wall
density field for $z<1$ arising from the {\em full} $\phi _{\mbox{\tiny R}0}$%
, we could then still use the linear response relation to determine the
large density change $\Delta \rho _{0,\mbox{\tiny 
R}}({\bf r})\equiv \rho _{0,\mbox{\tiny 
R}}({\bf r})-\rho _{0,\mbox{\tiny 
R}1}({\bf r})$ induced for $z>1$. Imposing accurate density values in
general is very difficult \cite{walloz}, but for the hard wall potential $%
\phi _{\mbox{\tiny 
R}0}=\phi _{HW}$ we have the exact result $\rho _{0,\mbox{\tiny 
R}}({\bf r})=0$ for all $z\leq 1$. Thus replacing $\delta \rho _{0,%
\mbox{\tiny 
R}1}$ by $\Delta \rho _{0,\mbox{\tiny 
R}}$ in (\ref{linresp}) and setting $\rho _{0,\mbox{\tiny 
R}}=0$ for all $z\leq 1$, we find for $z_1>1$: 
\begin{equation}
\Delta \rho _{0,\mbox{\tiny 
R}}({\bf r}_1)/\!\rho _{0,\mbox{\tiny 
R}1}({\bf r}_1)=\int \!d{\bf r}_2\,c_0({\bf r}_1,{\bf r}_2;{\bf [}\rho _{0,%
\mbox{\tiny 
R}1}])\Delta \rho _{0,\mbox{\tiny R}}({\bf r}_2).  \label{wallpy}
\end{equation}

Eq. (\ref{wallpy}) is a linear equation for $\Delta \rho _{0,\mbox{\tiny 
R}}({\bf r}_1),$ which we can directly solve by iteration or other means,
approximating $c_0({\bf r}_1,{\bf r}_2;{\bf [}\rho _{0,\mbox{\tiny 
R}1}])$ by that of an appropriate uniform system, just as we did before.
When $\rho _{0,\mbox{\tiny R}1}({\bf r)}=\rho _B,$ Eq. (\ref{wallpy}) is
equivalent to the usual hard-wall, hard-particle PY equation, which has an
analytic solution \cite{inteqns}. Eq. (\ref{wallpy}) is quite adequate for
our purposes here, though small errors can be seen at the highest densities.
If still more accuracy is required, we could use modified GMSA type
equations related to the PY equation \cite{inteqns}. It may also be possible
to use new and very accurate density functional methods for hard core fluids
in this step of our method \cite{rosenfeld}.

\begin{figure}[tbp]
\epsfxsize=3in
\centerline{\epsfbox{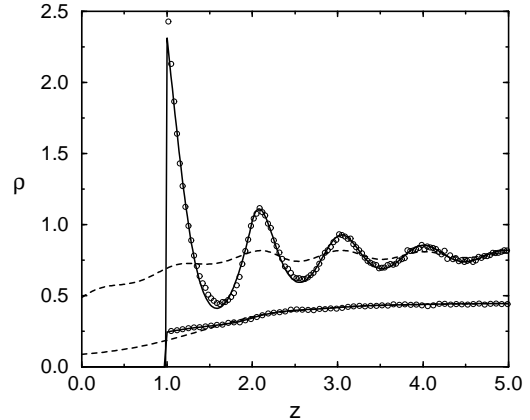}}
\caption{Density profiles $\rho _{0,\mbox{\tiny R}1}$ (dashed line) and $%
\rho _{0,\mbox{\tiny R}}$ (solid line) compared to MC simulations of the
reference fluid in potential $\phi _{\mbox{\tiny R}}$ (circles) for the same
states as in Fig.1.}
\end{figure}

The net result of this two step process is the desired $\rho _{0,%
\mbox{\tiny
R}}$ arising from a given $\phi _{\mbox{\tiny 
R}}.$ This can be substituted into Eq. (\ref{mfint}), which can then be
iterated to determine the final self-consistent $\phi _{\mbox{\tiny R}}.$ In
Fig. (1) we give the self-consistent potentials $\phi _{\mbox{\tiny R}%
1}=\phi _s$ that satisfy Eqs. (\ref{mfint}), (\ref{LMB}), and (\ref{wallpy})
for two different states along the near critical isotherm $T=1.35$. We see
that $\phi _{\mbox{\tiny R}1}$ is indeed a slowly varying repulsive
interaction in both cases. In Fig. (2) we give the associated smooth density
profiles $\rho _{0,\mbox{\tiny R}1}$ from (\ref{LMB}) for each state, as
well as the full profiles $\rho _{0,\mbox{\tiny R}}$ determined from Eq. (%
\ref{wallpy}). These are compared to Monte Carlo simulations we carried out 
\cite{ourotherpapers} of the reference system in the ERFs of Fig. (1). This
directly tests the accuracy of our two step procedure for calculating the
effects of $\phi _{\mbox{\tiny R}}$ on the reference system. The agreement
is excellent.

In Fig. (3) we test the simplified mean field treatment of the attractive
interactions in Eq. (\ref{mfint}) by comparing the reference profiles $\rho
_{0,\mbox{\tiny R}}$ to those of the full LJ fluid in the presence of the
hard wall, as determined by MC calculations. There is good agreement, though
small quantitative differences can be seen. Thus even the simplest mean
field treatment of attractive interactions is capable of capturing the major
changes in the density profile as the density is decreased, and at lower
temperature at coexistence we find complete drying states where a stable
vapor-liquid interface can exist arbitrarily far from the wall \cite
{ourotherpapers}.

Our emphasis thus far has been on quantitative numerical calculations.
However, the qualitative features of our method are equally important. A
long-standing problem of liquid state theory, well illustrated by the
nonuniform fluid example studied here, is how to treat consistently the
oscillating molecular scale ``excluded volume'' correlations arising from
repulsive intermolecular forces and the more slowly varying and longer
ranged correlations arising from attractive forces and often associated with
the formation of interfaces. In principle these issues are addressed by
modern density functional and integral equation methods, but in practice a
number of uncontrolled and often mathematically motivated approximation are
made. It is often difficult to assess their physical implications, and to
determine where the major sources of error lie. Here, we have divided this
problem into several distinct parts, whose accuracy can be examined
separately, and where the physical content and limitations of the methods
used are more clear. We used here the simplest mean field equation to
determine the ERF $\phi _{\mbox{\tiny R}}$ but more accurate (though more
complicated) equations derived from (\ref{exactybg}) are available \cite
{ourotherpapers}.

To determine the structure of the reference fluid in the presence of a given
ERF, we first calculated the response to $\phi _{\mbox{\tiny
R}1}$, the slowly varying part of the ERF, by expanding about a uniform
system. Next we used a gaussian field model \cite{crooks} (equivalent to a
modified wall-particle PY equation) to calculate the response to the
remaining harshly repulsive part $\phi _{\mbox{\tiny R}0}$ of the ERF. More
accurate methods could be used in both steps if necessary, and for
qualitative purposes both steps can be simplified considerably. For example,
Lum, Chandler, and Weeks \cite{lcw} have developed very simple
approximations for use with continuum Landau-Ginsburgh type equations that
give good qualitative results in a number of different cases, including
hydrophobic interactions in water. Application of these ideas to a variety
of different problems is underway \cite{ourotherpapers,lcw}.

This work was supported by NSF Grant No. CHE9528915. KV acknowledges support
from the Deutsche Forshungsgemeinschaft. We thank D. Chandler, K. Lum, and
J. Broughton for helpful discussions.

\begin{figure}[tbp]
\epsfxsize=3in
\centerline{\epsfbox{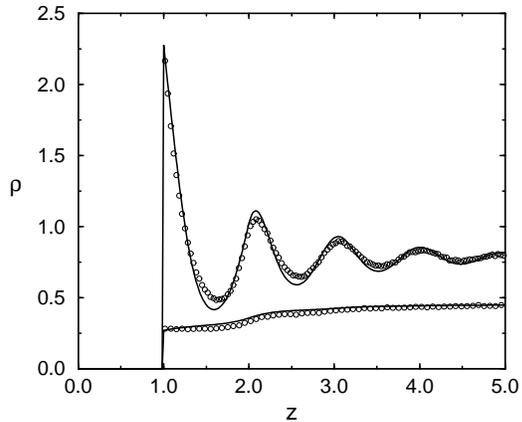}}
\caption{Density profiles $\rho _{0,\mbox{\tiny R}}$ compared to MC
simulation of the full LJ fluid (circles) for the same states as in Fig.1.}
\end{figure}


\begin{references}
\bibitem{walloz}  By definition \cite{frishleb}, the {\em exact} density
change $\Delta \rho _{0,\mbox{\tiny 
R}}({\bf r})\equiv \rho _{0,\mbox{\tiny 
R}}({\bf r})-\rho _{0,\mbox{\tiny 
R}1}({\bf r})$ that arises when $\phi _{\mbox{\tiny 
R}0}$ is taken into account satisfies Eq. (\ref{linresp}) when the factor $%
-\beta \delta \phi _{\mbox{\tiny 
R}1}({\bf r)}$ is replaced by the (wall-particle) {\em direct correlation
function} $C_{\mbox{\tiny 
R}0}({\bf r;[}\rho _{0,\mbox{\tiny 
R}1}]).$ Accurately determining this for a general potential is very
difficult. However, for the {\em hard wall}, if we make the the PY
assumption that $C_{\mbox{\tiny 
R}0}$ has the same range as $\phi _{\mbox{\tiny 
R}0}$ and hence vanishes for $z>1,$ we again arrive at Eq. (\ref{wallpy})
for $z>1.$
\end{references}
\end{document}